\documentstyle [preprint,aps,eqsecnum]{revtex}

\def\DA{[{\cal D} A]}
\def\Dpsi{[{\cal D} \psi]}
\def\Dpsib{[{\cal D} \bar\psi]}
\def\Df{[{\cal D} f]}
\def\ddk{{d^d k \over (2 \pi)^d}}

\def\bpsi{\bar \psi}
\def\x{|{\bf x}|}
\def\xy{|{\bf x}{-}{\bf y}|}
\def\E{{\rm E}}
\def\F{{\rm F}}
\def\R{{\rm R}}
\def\m{{\rm min}}
\def\g{{\rm gh}}
\def\A{{\rm A}}
\def\km{\sqrt{{\bf k}^2{+}m^2}}
\def\T{{\bf {\rm T}}}
\def\f{{\rm f}}
\def\mtop{m_{\rm topo}}

\begin{document}

\preprint{PURD-TH-94-12}

\title{Gauge-fixing parameter dependence of
	two-point gauge variant correlation functions}

\author{Chengxing Zhai}

\address{Department of Physics,
	 Purdue University,
	 West Lafayette, IN 47907}

\maketitle

\begin{abstract}
The gauge-fixing parameter $\xi$ dependence of two-point
gauge variant correlation functions
is studied for QED and QCD. We show that, in three Euclidean dimensions,
or for four-dimensional thermal gauge theories, the usual procedure of
getting a general covariant gauge-fixing term by averaging
over a class of covariant gauge-fixing conditions leads to a nontrivial
gauge-fixing parameter dependence in gauge variant two-point correlation
functions (e.g. fermion propagators). This nontrivial gauge-fixing
parameter dependence modifies the large distance behavior of the
two-point correlation functions by introducing additional exponentially
decaying factors. These factors are the origin of the gauge dependence
encountered in some perturbative evaluations of the damping rates and
the static chromoelectric screening length in a general covariant gauge.
To avoid this modification of the long distance behavior introduced
by performing the average over a class of covariant gauge-fixing conditions,
one can either choose a vanishing gauge-fixing parameter or apply
an unphysical infrared cutoff.
\end{abstract}

\newpage

\section{Introduction}
Physical observables such as thermal damping rates and the Debye screening
length which are determined by the position of poles (or generally,
singularities) in correlation functions are gauge invariant
quantities\cite{Braaten&Pisarski,%
Kobes&Kunstatter&Rebhan}. However, for a general covariant
gauge, a gauge-fixing parameter $\xi$ dependence in
damping rate was reported\cite{Baier&Kunstatter&Schiff} in a
perturbative evaluation of the fermion damping rate.
Similar gauge-fixing parameter dependence was also
encountered in the perturbative calculation of the Debye screening
length at the next-to-leading order\cite{Rebhan1}.
A way to extract the gauge independent damping rate and screening length
is to introduce an unphysical infrared cutoff\cite{Rebhan2}.
In this article, we examine
the gauge-fixing parameter $\xi$ dependence of two-point correlation
functions. As is well known, a conventional way of getting a
general covariant gauge-fixing term $(\partial A)^2/(2\xi)$ involves
performing an average over a class of covariant gauge conditions.
We shall show that, in three Euclidean dimensions or for four-dimensional
thermal gauge field theories, this average over different gauge conditions
generates a nontrivial $\xi$ dependence in two-point correlation functions of
gauge variant operators such as the fermion propagator. Specifically, for
QED and to leading order of QCD, this gauge-fixing parameter dependence
alters the long range behavior of the two-point correlation functions
by an extra exponentially decaying factor with the exponent depending
on $\xi$. This is the origin of the $\xi$ dependence encountered in
perturbative calculations of the damping rate and the Debye screening
length\cite{Baier&Kunstatter&Schiff,Rebhan1,Rebhan2}.
This gauge-fixing parameter $\xi$ dependence is an artificial
fact due to doing the average over a class of gauge conditions
since this average includes gauge conditions $\partial A = f$ with
$f$ containing long wavelength fluctuations.
Choosing the Landau gauge $\xi=0$ which means taking the ``no average'' limit
removes the modification of the long distance behavior produced by averaging
over gauge conditions.
Another way is to introduce an unphysical infrared cutoff as
suggested in reference\cite{Rebhan2} to suppress the contributions
to the gauge condition average from the infrared fluctuations.
If the physical content of the theory is gauge invariant in the sense
that the gauge constrain $\partial A = 0$ is equivalent to other
physical gauge constrains when quantizing the theory, choosing $\xi=0$
then yields the correct physics ({\it e.~g.} correlation length).

In next section, we shall first study generally how the average over
a class of gauge choices produces the $\xi$ dependence of the propagators
of charged particles in QED where it is completely solvable.
The $\xi$ dependence of the propagator of a charged particle
may be expressed simply as a $\xi$-dependent multiplicative factor
which becomes an exponentially decaying factor for large spacetime
argument in three Euclidean dimensions or four-dimensional thermal QED.
Thus, if one extracts the correlation length or the damping rate from
the propagator of a charged particle, a $\xi$-dependent correlation length
or damping rate is obtained. We then explain why an unphysical infrared
cutoff can get rid of this $\xi$ dependence and produce the same
result as the choice $\xi=0$.
In section III, we perform parallel analysis for QCD. Since a complete
solution could not be achieved, we only do a leading order perturbative
calculation. To the leading order, the gauge-fixing parameter dependence
for the fermionic propagator is the same as in QED with an effective charge.
The $\xi$ dependence of the static (chromo)electric screening length is
discussed. We draw conclusions in section IV. In Appendix A, details
about the evaluation of a thermal integral is given. In Appendix B,
we present another way of deriving the $\xi$ dependence in fermion propagators
based on Ward identities for the proper vertices.

\section{$\xi$ dependence of a charged particle propagator in QED}

\subsection{A functional derivation}
In this subsection, we shall derive the gauge-fixing parameter $\xi$
dependence of the propagator of a charged particle in QED in a
general covariant gauge. Since the derivation
does not depend on whether the charged particle is a scalar or
a fermion, we focus on the fermion case.
Throughout this paper, we work in Euclidean spacetime.
Real time results can then be obtained by analytic continuation.

The fermion propagator\footnote{We use $\langle \cdots \rangle$
to represent the Euclidean time ordered product.}%
is defined by the Euclidean functional integral as
\begin{eqnarray}
    G_f (x, y) &\equiv& \langle \bar\psi (x)
	\psi (y) \rangle
\nonumber\\
	&\equiv& \int \DA \Dpsib \Dpsi \exp \left \{
	- \int {\cal L}_\E \right \} \bar\psi (x)
	\psi (y) \delta (\partial A - f) \,,
\label{greendef}
\end{eqnarray}
where we have introduced the gauge-fixing condition
\begin{equation}
    \partial A (x) = f (x)
\end{equation}
with $f (x)$ being an arbitrary function.
Denoting the charge of the fermion by $q$,
it is not hard to justify that the gauge transformation
change of variables
\begin{eqnarray}
    \psi (x) &\to& \psi' (x) = e^{i q \lambda (x)} \psi (x)
\nonumber\\
    \bar\psi (x) &\to& \bar\psi' (x) = e^{-i q \lambda (x)} \bar\psi (x)
\nonumber\\
    A (x) &\to& A' (x) = A (x) - \partial \lambda (x) \;,
\end{eqnarray}
with $\lambda$ satisfying
\begin{equation}
    \partial^2 \lambda (x) = f (x) \,,
\end{equation}
changes the gauge-fixing condition
\begin{equation}
    \partial A = f \qquad \to \quad \partial A' = 0
\end{equation}
and thus gives
\begin{equation}
    G_f (x, y) = G_{f=0} (x, y) \, e^{i q (\lambda (x) - \lambda (y))} \,.
\end{equation}
Defining Green's function $\Delta (x-x')$ by
\begin{equation}
    \partial_x^2 \, \Delta(x-x') = \delta (x-x')
\end{equation}
enables us to write
\begin{equation}
    \lambda (x) = \int dz \, \Delta (x-z) f (z) \,.
\end{equation}
The usual procedure of getting the general covariant gauge now involves
averaging over $f (x)$ with a weighting factor $\exp \{-{1 \over 2\xi}
\int dx f^2 \}$. Let us denote the fermion propagator in a general
covariant gauge by $G_\xi (x, y)$. Then
\begin{eqnarray}
    G_\xi (x, y) &=& \int \Df G_f (x, y) \exp \left \{
	- {1 \over 2 \xi} \int dz f^2 (z) \right \}
\nonumber\\
	&=& \int \Df \exp \left \{ - \int dz \left [
	{1 \over 2 \xi}  f^2 (z) + i J (z;x,y) f(z) \right ]
	\right \} G_{f=0} (x, y) \,,
\label{gaussian}
\end{eqnarray}
where we have defined the linear source $J (z;x,y)$ as
\begin{equation}
    J (z;x,y) = - q \, \left [\Delta (z-x) - \Delta (z-y) \right ] \,.
\label{source}
\end{equation}
A straight forward evaluation of the gaussian functional integral
in Eq.~(\ref{gaussian}) yields
\begin{equation}
    G_\xi (x, y) = \exp \left \{ - {\xi q^2 \over 2}
	\int dz \, \left [\Delta(z-x) - \Delta(z-y) \right ]^2
	 \right \} G_{f=0} (x, y) \,,
\label{QEDresult}
\end{equation}
where the normalization factor for the functional integral has been
chosen so that
\begin{equation}
    \int \Df \exp \left [ - {1 \over 2 \xi}
	\int dz \, f^2 (z) \right ] = 1 \,.
\end{equation}
It is trivial to check that the Landau gauge choice $\xi = 0$
corresponds to the gauge choice $f = 0$.
This is expected since the weighting functional
$\exp \{-{1 \over 2\xi} \int dx f^2 \}$ for the average
allows $f$ to fluctuate around $f=0$ with the variance proportional
to $\xi$. Setting $\xi=0$ confines $f$ to be $0$.
The gauge-fixing parameter dependence appears as a multiplicative factor.
Result~(\ref{QEDresult}) is also valid for the
propagator of a scalar charged particle with charge $q$.
Equation~(\ref{QEDresult}) has been derived by other methods and discussed
for four-dimensional QED\cite{Brown,Zinn-Justin}.

By including the photon source and more pairs of the
Green's function $\Delta(x)$ in the source~(\ref{source}),
Eq.~(\ref{QEDresult}) may be generalized to cases where the correlation
functions contain external photon lines and additional fermion lines.
The $f$ functional integral is still a gaussian integral.
We omit the algebraically complicated intermediate steps which are
completely parallel to those for the fermion propagator.
The $\xi$ dependence for a correlation function with
$2n$ external fermion legs is
\begin{eqnarray}
    &&\langle \bpsi (x_1) \bpsi (x_2) \cdots \bpsi (x_n) \,
	 \psi (y_1) \psi (y_2) \cdots  \psi (y_n) \,
	e^{i q \int j \cdot A} \rangle_{\xi}
\nonumber\\
	&& \qquad = \exp \left \{ - {\xi q^2 \over 2}
	\int dz \left [
	\sum_{i=1}^n \left (\Delta(z{-}x_i) - \Delta(z{-}y_i) \right )
	+ \int dz' \Delta (z{-}z') \,
	\partial_{z'}{\cdot} j (z') \right ]^2 \right \}
\nonumber\\
	&& \quad \qquad
	\times \langle \bpsi (x_1) \bpsi (x_2) \cdots \bpsi (x_n) \,
         \psi (y_1) \psi (y_2) \cdots  \psi (y_n) \,
        e^{i q \int j \cdot A} \rangle_{f=0} \,,
\label{2n+photon}
\end{eqnarray}
where we have introduced the photon source term $ i q \int j \cdot A$.
Taking derivatives with respect to the photon source $j$ gives
insertions of the photon fields in the correlation function.
We note that the $\xi$ dependence is totally factorized. Therefore,
the choice $\xi=0$ is equivalent to the gauge choice $f=0$.

\subsection{$\xi$ dependence in four and three dimensional QED}
We now study the behavior of the multiplicative $\xi$-dependent factor
in four and three dimension spacetime.
We do not consider the case $d=2$ since there infrared divergences
are so serious that the charged particles are confined\cite{Schwinger}.
To facilitate the notation, let us define
\begin{equation}
    I (x) \equiv {1 \over 2} \int dz \,
	\left [\Delta(x-z) - \Delta(z) \right ]^2
\label{I_def}
\end{equation}
so that
\begin{equation}
    G_\xi (x, y) = e^{- \xi q^2 I (x-y)} G_{f=0} (x, y) \,.
\label{QEDresult2}
\end{equation}
In momentum space, $I(x)$ can be expressed\footnote{Here, we have
made a change of variable $k \to -k$. In three dimensions, this
causes an infrared problem which does not really matter if we
only use regulators invariant under the inversion of momentum.} as
\begin{equation}
    I (x) = {1 \over 2}
	\int (dk) \, {1 \over k^4} (2 - e^{i k x} - e^{-i k x})
	= \int (dk) {1 \over k^4} (1 - e^{i k x}) \,,
\label{I_mom_exp}
\end{equation}
where $\int (dk)$ represents the appropriate momentum integral
conjugate to the spacetime.
In $d$ dimensional Euclidean spacetime,
\begin{equation}
    I (x) = \int \ddk {1 \over k^4} \, (1 - e^{i k x}) \,,
\end{equation}
which can be evaluated as
\begin{eqnarray}
    I (x) &=& \int_0^\infty ds \, s \int \ddk \, e^{-s k^2} (1 - e^{i k x})
\nonumber\\
	&=& {1 \over (4 \pi)^{d/2}} \int_0^\infty ds \, s^{1-d/2}
	\exp \left [- {x^2 \over 4 s} \right ]
\nonumber\\
	&=& - {\Gamma(d/2 - 2) \over (4 \pi)^2}
	\left (\pi x^2 \right )^{2-d/2} \,.
\label{Schwinger}
\end{eqnarray}
Setting $d=4$, Eq.~(\ref{QEDresult2}) takes the form
\begin{equation}
    G_\xi (x, y) = \exp \left \{ {\xi_\R q_\R^2 \over (4 \pi)^2}
	{2 \over d - 4} \right \}
	\left [\pi\mu^2 (x - y)^2 \right ]^{-\xi_\R q_\R^2/(4 \pi)^2}
	e^{- \xi_\R q_\R^2 \gamma /(4 \pi)^2} \, G_{f=0} (x, y)
	+ O(d-4) \,,
\label{4dQED}
\end{equation}
where we have introduced the renormalization scale $\mu$ by
\begin{equation}
    q^2 \, \xi = q_\R^2 \, \xi_\R \, \mu^{4-d}
\end{equation}
with $q_\R^2$ and $\xi_\R$ being the renormalized charge and gauge-fixing
parameter. $\gamma$ in Eq.~(\ref{4dQED}) is the Euler constant.
The ultraviolet divergence appearing in the gauge dependent
factor gives the usual gauge dependence of the fermion wave
function renormalization factor in agreement with previous
results\cite{Brown,Collins,Lautrup} as is
the $(x-y)^2$ power modification factor\cite{Brown}.
In four dimensions, the $\xi$ dependence in the fermion propagator
modifies the long distance power law behavior. Due to massless photons,
the fermion propagator does not exhibit a pole in momentum space
but a branch cut with the behavior
$(p^2 + m^2)^{-(1{+}\nu)}$\cite{Zwanziger,Abrikosov&Gorkov&Svidzinskii}.
The $\xi$-dependent modification factor~(\ref{4dQED})
leads to a $\xi$-dependent $\nu$ and thus a $\xi$-dependent
on-shell condition for fermions in four dimensional QED\cite{Brown}.

We now consider QED in d=3 Euclidean space%
\footnote{For a real three dimensional QED, there
is evidence\cite{Morchio&Strocchi,Sen} showing that the
charged particles may be confined in three dimension QED
due to infrared divergences. Of course, one can consistently
add to the theory a topological mass term for the photon field
without breaking the local gauge symmetry\cite{Jackiw,Schonfeld}.
This topological mass term can be even dynamically generated by
interacting with the fermions
\cite{Jackiw,Schonfeld,Niemi&Semenoff,Redlich}. With this
topological mass term, the charged particles are no longer
confined\cite{Sen,Jackiw}.}. We can view this
as a dimensionally reduced field theory of a four dimensional
scalar QED at the high temperature limit\cite{Gross&Pisarski&Yaffe,%
Appelquist&Pisarski}. $I (x)$ is both ultraviolet and infrared finite.
Eq.~(\ref{Schwinger}) reads explicitly $I ({\bf x}) = \x /(8 \pi)$.
We have then
\begin{equation}
    G_\xi ({\bf x}, {\bf y}) =
	e^{- {\xi q^2 \over 8 \pi} \xy} \,
	G_{f=0} ({\bf x}, {\bf y}) \,.
\label{QED3d}
\end{equation}
Hence the usual average over covariant gauge
conditions has introduced a $\xi$-dependent exponentially decaying
factor to the propagators of charged particles. Thus, the correlation
length also acquires $\xi$ dependence. If we extract the correlation length
from a gauge variant two-point correlation function,
different $\xi$ values give different answers.
Choosing $\xi = 0$ gives the answer corresponding to the original gauge
theory quantized by the gauge condition\footnote{%
It should be mentioned that if we view the three dimension theory as
the high temperature limit of a four dimension theory,
the gauge condition $\partial A = 0$ really corresponds to
the Coulomb gauge choice in the original four dimensional theory.
} $\partial A=0$. This justifies a previous claim on
the preference of the Landau gauge choice\cite{Jackiw}.
We note here that if only the gauge invariant correlation functions
are considered, there are no $\xi$ dependences inside the correlation
functions which states that all $\xi$'s are equivalent for
gauge invariant correlation functions.

We now examine the origin of the decaying factor in Eq.~(\ref{QED3d})
and explain why an unphysical infrared cutoff can remove this
artificial $\xi$-dependence.
In previous subsection, we found the relation
\begin{equation}
    G_f (x, y) = \exp \left \{-i q \int dz \,
	\left [ \Delta (x-z) - \Delta (y-z) \right ] f(z) \right \}
	G_{f=0} (x, y)
\end{equation}
which basically states that different choices of $f$ are equivalent.
However, this statement is based on the assumption that the factor
\begin{equation}
    \exp \left \{-i q \int dz \,
	\left [ \Delta (x-z) - \Delta (y-z) \right ] f(z) \right \}
\label{factor}
\end{equation}
does not alter the long distance behavior of the propagator
or, in particular, the position of the
physical pole appearing in the Fourier transform of $G_f (x, y)$.
If $f(z)$ is a localized function so that
the exponent in factor~(\ref{factor}) vanishes as $x$ or $y$
becomes large, $G_f (x, y)$ has the same large distance
behavior as $G_{f=0} (x, y)$. We can then conclude that the
gauge choice $\partial A = f$ is equivalent to the choice
$\partial A = 0$.
However, when performing the average of gauge conditions,
nonlocalized $f$'s are not excluded.
It is not hard to see that
the weighting functional\footnote{Here ${\tilde f} (k)$
is the Fourier transform of $f(z)$.}%
\begin{equation}
    \exp \left \{ - {1 \over 2 \xi} \int dz f^2 (z) \right \}
	= \exp \left \{ - {1 \over 2 \xi}
	\int \ddk \, {\tilde f} (k) \, {\tilde f} (-k) \right \}
\end{equation}
contains long wavelength modes (${\tilde f} (k)$ with
wavelength $1/k$ longer than the separation between
$x$ and $y$). It is the inclusion of these long wave modes in the
average that yields an exponentially decaying factor depending on
$\xy$.
For the gauge conditions with $f$ having wavelength shorter than
$\xy$, the $x, y$ dependence in phase factor~(\ref{factor})
is washed out after summing over different short wavelength contributions.
In another word, the short wave fluctuations do not suffice to change
the behavior of the long range correlation after the average.
Indeed, $I(x)$ does get its main contribution from the infrared
region with $k$ being order $1/\x$ or less. This generates a piece
proportional to $\x$ and therefore results an exponentially
decaying factor. Employing an unphysical infrared cutoff to suppress the
contributions from these long wave $f$'s can eliminate the
gauge-fixing parameter dependence in the correlation length.
Explicitly, introducing an infrared cutoff $k_\m$ to integral $I ({\bf x})$,
we have, as $\x \to \infty$,
\begin{eqnarray}
    \int_{k>k_{\rm min}} {d^3k \over (2 \pi)^3}
	{1 \over k^4} \, (1-e^{i {\bf k} \cdot {\bf x}})
	&=& {1 \over 2 \pi^2} \int_{k_\m}^{\infty} dk
	{1 \over k^2} \left (1 - {\sin k \x \over k \x} \right )
\nonumber\\
	&=& {1 \over 2 \pi^2} \left ({1 \over k_\m}
	- \x \int_{k_\m \x}^\infty ds {\sin s \over s^3} \right )
	\to {1 \over 2 \pi^2 k_\m}
\end{eqnarray}
which does not depend on ${\bf x}$.
Therefore, the average over the short wave $f$'s does not modify the
long distance behavior of the propagators but an overall constant.
This explains why an unphysical infrared cutoff proposed in
reference\cite{Rebhan1,Rebhan2} can remove the $\xi$-dependent
modification of the long distance behavior.

Since lower dimension field theories are more sensitive to the infrared
region, this exponentially decaying factor does not appear
in $d=4$. In four dimensions, only the power law of the propagator
is changed. For four-dimensional thermal field theories,
the imaginary time formalism
leads to a Euclidean functional representation for thermal
correlation functions with one dimension of the spacetime compactified.
Therefore, we expect that the average over different covariant gauge choices
may also cause serious modifications to the two-point correlation functions.

\subsection{Four-dimensional thermal QED}
We now study the four-dimensional thermal QED which is equivalent to
QED in three spatial dimensions plus an additional compactified
imaginary time dimension. As such,
the momentum in this dimension is discretized so that
the momentum integral for $I (x)$ in Eq.~(\ref{I_mom_exp}) is
a ``sum-integral'':
\begin{equation}
    \int (dk) \to T \sum_{k_0} \int
	{d^{d-1} k \over (2 \pi)^{d-1}} \,.
\end{equation}
Since there is the zero temperature part contributing to the sum-integral,
it is ultraviolet divergent. Using the dimensional regularization, we
calculate this sum-integral in Appendix A and simply
quote the result~(\ref{sumresult}) as:
\begin{eqnarray}
    I(\tau, {\bf x}) &=& {1 \over (4 \pi)^2} (4 \pi T^2)^{d/2-2}
	\left [{2 \over 4-d} + \gamma \right ]
\nonumber\\
	&& \qquad + {1 \over (4 \pi)^2} \ln \left (1 + e^{-4 \pi T \x}
	-2 \cos (2 \pi T \tau) e^{-2 \pi T \x} \right )
	+ {T \x \over 8 \pi} + O (d-4) \,.
\end{eqnarray}
Here $I(x)$ has been written as $I(\tau, {\bf x})$ with $\tau$ and ${\bf x}$
being the imaginary time and spatial coordinate respectively.
It is not hard to check that for large $\x$ or high $T$
\begin{equation}
    I (\tau, {\bf x}) \to { T \x \over 8 \pi}
\end{equation}
which is in agreement with the analysis for the three dimensional Euclidean
theory discussed in previous section.
Therefore, the correlation length contains a $\xi$-dependent piece
$\xi q^2 T \x / (8 \pi)$.

So far the correlation functions studied are all defined in imaginary time.
To study the damping rate, a retarded real-time correlation function
is required.
We can analytically continuate it into real-time by replacing $\tau$
with $i t$ to obtain the corresponding real-time propagator.\footnote{This
analytic continuation yields real-time ordered correlation functions. However,
it is not hard to show that the exponentially decaying factor we
are concerned is the same as that for the retarded correlation function.}
Performing this analytic continuation for $I(\tau, {\bf x})$ gives
\begin{eqnarray}
    I(\tau, {\bf x}) &=& {1 \over (4 \pi)^2} (4 \pi T^2)^{d/2-2}
	\left [{2 \over 4-d} + \gamma \right ]
\nonumber\\
	&& \qquad + {1 \over (4 \pi)^2} \ln \left (1 + e^{-4 \pi T \x}
	-2 \cosh (2 \pi T t) e^{-2 \pi T \x} \right )
	+ {T \x \over 8 \pi} + O (d-4) \,.
\end{eqnarray}
For large $\x$ or $|t|$, we have
\begin{equation}
    I (t, {\bf x}) \sim {T \over 8 \pi} \left [
	|t| \, \theta (|t| - \x) + \x \theta (\x - |t|) \right ] \,.
\end{equation}
Inserting this large coordinate argument behavior
into Eq.~(\ref{QEDresult2}) gives
\begin{equation}
    G_\xi (t, {\bf x}) \sim
	\exp \left \{- {\xi_\R q_\R^2 T \over 8 \pi} \left [
	|t| \, \theta(|t|-\x) + \x \theta (\x - |t|) \right ] \right \}
	G_{f=0} (t, {\bf x}) \,,
\end{equation}
where we have shortened\footnote{We have implicitly switched back and forth
between the notations $x$ and $(t,{\bf x})$ for the spacetime coordinate.}
the Green's function notation $G (x, y)$ to $G (x{-}y)$
because of the spacetime translation invariance.
The $\xi$-dependent large time damping factor causes a
$\xi$-dependent damping rate if we use the general covariant gauge.
To avoid any modification of the large time behavior coming from
averaging the gauge conditions, we can choose $\xi=0$.
This choice yields the damping rate for the original gauge theory,
if the theory is invariant for different gauge constrains
(without involving any average) when quantizing it.
We also provide a more familiar derivation of Eq.~(\ref{QEDresult})
based on the proper vertex Ward identities in Appendix B where
Equation~(\ref{compare}) shows that the
$\xi$ dependence we discussed above is the same $\xi$ dependence
as reported in reference\cite{Baier&Kunstatter&Schiff}.
We now examine again why an unphysical
infrared cutoff can also remove this $\xi$ dependence\cite{Rebhan2}.
Introducing a small mass term $m^2$ to cutoff
the $k$ integral for $I(\tau,{\bf x})$ at the infrared region,
and using the usual contour trick to do the $k_0$ sum,
the finite temperature part of $I (\tau, x)$ is expressed as
\begin{equation}
    I^{(T)} (\tau, {\bf x}) = - {d \over d m^2} \left \{
	\int {d^3 k \over (2 \pi)^3} {n \, (\km) \over 2 \km}
	\left [ e^{i{\bf k}\cdot{\bf x}
	- \tau \km} + e^{i{\bf k}\cdot{\bf x}
	+ \tau \km} - 2 \right ] \right \} \,,
\end{equation}
where $n(\omega)$ is the Bose distribution factor
\begin{equation}
    n (\omega) \equiv {1 \over e^{\beta \omega} - 1}
\end{equation}
with $\beta = 1/T$.
Doing the analytic continuation $\tau \to it$, we find that
as $t$ or $\x$ becomes large, the Riemann-Lebesgue lemma kills
$t$ and $\x$-dependent part in $I^{(T)} (t, {\bf x})$. Hence the large
coordinate argument behavior does not obtain any $\xi$ dependence
but an overall constant.
This shows why an unphysical cutoff in reference\cite{Rebhan2} can
remove the gauge-fixing parameter dependence in the damping rate.

\section{Gauge-fixing parameter dependence in thermal QCD}
The gauge-fixing parameter dependence we studied for QED can also be derived
by using Ward identities.
Since to leading order in QCD, Ward identities are the same
as that in QED, we expect that at the leading order, we should be
able to find similar gauge-fixing parameter dependence as occurs
in QED. This is indeed the case. Since the derivation for QED is
completely parallel to the leading order derivation for QCD,
it suffices to just provide the derivation for QCD at the leading
order.
\subsection{fermionic damping rate}
Let us consider the fermion propagators. To simplify the notation,
we shall suppress the color indices of the fermion fields.
We shall use the Ward identities to derive the result for the QCD
fermion propagator analogous to the result~(\ref{QEDresult}). Taking the
derivative with respect to $\xi$ which introduces an insertion
of the gauge-fixing term in the functional integral,
we obtain
\begin{equation}
    {d \over d \xi} G_\xi (x, y) = {1 \over 2 \xi^2}
	\int dz \, \langle \, \bar\psi (x) \psi (y)
	\left [\partial A^a (z) \right ]^2 \, \rangle \,,
\end{equation}
where $\psi$ and $A$ represent the fermion fields and the
gauge fields respectively
and $a$ is the color index in the adjoint representation.
Using the equation of motion for the ghost field, the
BRS transform of the correlation function
$\langle \bar\psi(x) \psi (y) \partial A^a (z)
\bar c^a (z) \rangle $ produces the relation
\begin{equation}
    \langle \, \bar\psi(x) \psi(y)
	\left [\partial A^a (z) \right ]^2 \rangle
	= i g \, \xi \, \langle \bar\psi(x)
	\left [c^b (x) - c^b (y) \right ]
	\T^b \psi (y) \left [\partial A^a (z) \right ]
	\bar c^a (z) \, \rangle \,,
\end{equation}
where $\T^b$ are the generators in the fermion representation and
$\bar c^b (z)$ and $c^b (z)$ are the Faddeev-Popov ghost fields. Noting that
\begin{equation}
    \langle c^b (x) A^a (y) \bar c^a (z) \rangle = 0 \,,
\end{equation}
we have, at the leading order,
\begin{eqnarray}
    \langle \bar\psi(x) \psi(y) \left [\partial A^a(z) \right ]^2 \rangle
	&=& \! {-} i g \, \xi \, \langle \bar\psi(x) \T^b \psi(y)
	\left [\partial A^a(z) \right ] \rangle \,
	\left [\langle c^b (x) \bar c^a (z) \rangle -
	\langle c^b (y) \bar c^a (z) \rangle \right ] + O (g^4)
\nonumber\\
	&=& \! {-}i g \, \xi \,
	\langle \bar\psi(x) \T^a\psi(y) \left [\partial A^a(z) \right ]
	\rangle \, \left [\Delta_\g (x{-}z){-}\Delta_\g (y{-}z) \right ]
	 + O (g^4) \,,
\label{QCD_WI1}
\end{eqnarray}
where the ghost propagator $\Delta_\g (x)$ is defined as
\begin{equation}
    \langle c^a (x) \bar c^b (y) \rangle \equiv \delta^{ab}
	\Delta_\g (x-y) \,.
\end{equation}
Similarly, the BRS transform of
$\langle \bar\psi(x) \T^a \psi (y) \bar c^a (z) \rangle$ yields
\begin{eqnarray}
    \langle \, \bar\psi(x) \T^a \psi(y)
	\left [\partial A^a (z) \right ] \, \rangle
	&=& - i g \, \xi \, \langle \bar\psi(x) \T^a\T^a \psi (y) \rangle \,
	\left [\Delta_\g (x{-}z) - \Delta_\g (y{-}z) \right ] + O (g^3)
\nonumber\\
	&=& - i g \, \xi \, C_\F \, G_\xi (x, y) \,
	\left [\Delta_\g (x{-}z) - \Delta_\g (y{-}z) \right ] + O (g^3) \,,
\label{QCD_WI2}
\end{eqnarray}
where $C_\F$ is the Casimir for the fermion representation.
Combining results~(\ref{QCD_WI1}) and (\ref{QCD_WI2}) above gives
\begin{equation}
    {d \over d \xi} G_\xi (x, y)
	\simeq - {1 \over 2} g^2 C_\F \int dz \,
	\left [\Delta_\g (x{-}z) - \Delta_\g (y{-}z) \right ]^2 \,
	G_\xi (x, y) + O (g^4) \,.
\end{equation}
Perturbatively,
\begin{equation}
    \Delta_\g (x) = \Delta (x) + O (g^2).
\end{equation}
Therefore, to the leading order, the above
differential equation shows that the leading $\xi$ dependence of
the fermion propagator can be expressed as
\begin{equation}
    G_\xi (x, y) \simeq \exp \left \{
	- {\xi g^2 C_\F \over 2} I \, (x{-}y)
	\right \} G_{f=0} (x, y) \,.
\label{QCDresult}
\end{equation}
We note here that a QED derivation may be obtained simply by
ignoring all the color indices, changing $\T^a \to 1$, $g \to q$,
and setting all the leading approximation to be exact.
For QCD, the leading\footnote{This is also true after performing the
Braaten-Pisarski resummation
since the resumed proper vertices and propagators still satisfy
the QED-type Ward identities\cite{Braaten&Pisarski,Taylor}.}
gauge-fixing parameter dependence of the fermion propagator
is the same as the exact $\xi$ dependence of a fermion propagator
in QED with effective charge $g C_\F^{1/2}$.
As discussed before, at finite temperature, if we extract the damping rate of
fermionic excitations from the fermion propagator calculated in
a general covariant gauge, we shall get a $\xi$-dependent damping rate.
This is the dependence reported in reference \cite{Baier&Kunstatter&Schiff}.
We can either choose Landau gauge $\xi = 0$ or use an unphysical
infrared cutoff to get rid of the modification of the long time
behavior due to taking the average over gauge conditions.

\subsection{Static electric screening length}
We now turn to consider the static chromoelectric screening length
to next-to-leading order.
Since the relevant energy scale is $gT$, it is convenient to
use the dimensionally reduced effective theory (so called EQCD)
which involves only the static gauge fields to describe thermal
QCD\cite{Nadkarni}. To the order we are concerned, we only need to study
a three dimensional Euclidean effective theory with Lagrangian
\begin{equation}
    {\cal L}_{\rm EQCD} = {1 \over 4} F_{ij}^a F_{ij}^a
	+ {1 \over 2} (D_i A_0)^a (D_i A_0)^a
	+ {1 \over 2} m^2_{\rm el} A_0^a A_0^a
\label{EQCD}
\end{equation}
where the covariant derivative $D_i$ is defined as
\begin{equation}
    (D_i A_0)^a  = \partial_i A_0^a - i g f^{abc} A_i^b A_0^c
\end{equation}
and $m_{\rm el}$ is the leading order Debye mass with the value
\begin{equation}
    m^2_{\rm el} = {1 \over 3} \left (C_\A + {N_{\rm f} \over 2} \right )
	g^2 T^2 \,.
\end{equation}
Here, $C_\A$ is the Casimir of the adjoint representation for the
gauge group and $N_\f$ is the number of fermion flavors.
Of course, we shall study, for a general covariant gauge, the
gauge-fixing parameter $\xi$ dependence of the static propagator
$D^{ab} ({\bf x},{\bf y})$ of the $A_0^a$ fields.
We have not explicitly included the
gauge-fixing term nor the ghost fields term. We employ
the same steps as in the last subsection for calculating
the gauge dependence of a fermion propagator.
All we need to do
is to change all the fermion fields into the $A_0^a$ fields and replace
the generators $\T^a$ by the generators in the adjoint
representation under which $A_0^a$ transforms.
Thus, the effective charge squared $g^2 C_\F$ appearing
in Eq.~(\ref{QCDresult}) is changed to $g^2 C_\A$.
After these replacements, we obtain
\begin{equation}
    D^{ab}_\xi ({\bf x},{\bf y}) \simeq
	\exp \left \{ - {\xi g^2 C_\A \over 8 \pi}
	|{\bf x} - {\bf y}| \right \}
	D^{ab}_{f=0} ({\bf x}, {\bf y}) \,.
\end{equation}
Performing the Fourier transform gives again a $\xi$-dependent singularity
in the propagator. This dependence appears in
the result\footnote{There, a $\xi$-dependent term
$\xi g^2 N m_{\rm el} T / (4 \pi)$ contributes to the self energy
which indicates a correction $\xi g^2 N T / (8 \pi)$
to the leading screening mass $m_{\rm el}$. $N$ is the number
of colors, or equivalently, the Casimir of the adjoint representation
for SU(N) gauge group.}
found in references\cite{Rebhan1,Rebhan2} where the
$\xi$ dependence is removed by introducing an unphysical cutoff
at the infrared region of the loop integral involved in
the self energy evaluation.

To avoid confusion, we like to add following comment.
It appears that
in reference\cite{Rebhan1}, the Feynman gauge choice, $\xi=1$,
coincides with the result obtained by introducing an unphysical
infrared cutoff. The free gauge boson propagator
\begin{equation}
    G (p) = {\delta_{ij} - \hat p_i \hat p_j \over p^2 } +
	\xi { \hat p_i \hat p_j \over p^2}
\end{equation}
contains the combination $\xi-1$ as the coefficient of
$\hat p_i \hat p_j/p^2$.
At the one-loop order, it can be shown explicitly\cite{Rebhan1}
that the part in the gauge boson propagator proportional to
$\hat p_i \hat p_j/p^2$
contributes to the self energy and causes a shift of the position of
the pole of $A_0$ propagator. Introducing a technical infrared cutoff
removes this contribution to the pole position of this propagator
from the $\hat p_i \hat p_j /p^2$ part of the gauge boson propagator.
This can be mimicked by choosing $\xi=1$.
However, the transverse part of the gauge boson propagator receives
additional radiative corrections while the pure longitudinal part does not
as a consequence of the Ward identity.
Due to these radiative corrections, this transverse piece
becomes less singular at the infrared region\cite{Rebhan1,%
Appelquist&Pisarski} and a {\it physical} infrared cutoff gets induced.
Consequently, its $\hat p_i \hat p_j$ piece
does not shift the position of the pole in $A_0$ propagator.
On the other hand, the longitudinal part remains the same and
does shift the position of pole except in Landau gauge with $\xi=0$.
Thus, the Landau gauge choice produces the same result as
a naive choice of Feynman gauge where in addition the radiative corrections
to the gauge boson propagator are ignored.
For theories containing a topological mass term, the gauge boson
propagator has the form
\begin{equation}
    G (p) = {\delta_{ij} - \hat p_i \hat p_j \over p^2 + \mtop^2}
	+ {\mtop \, \epsilon_{ijk} \, p_k \over (p^2 + \mtop^2) \, p^2}
		+ \xi {\hat p_i \hat p_j \over p^2}
\end{equation}
where $\mtop$ is the topological mass. Here it is easy to see that the
infrared behavior of the longitudinal part is different from that of the
transverse part. Choosing a vanishing gauge-fixing parameter is
the same as putting an infrared cutoff.

\section{Conclusions}
In conclusion, we have shown that in three Euclidean dimensions
or for four-dimensional thermal gauge field theories,
the usual averaging procedure for getting a general covariant gauge-fixing
term may introduce gauge-fixing parameter dependent modifications
to the large distance behavior of the gauge dependent correlation
functions. The gauge-fixing parameter dependent modification to
the large distance
behavior of the correlation function is the origin of the gauge
dependence encountered in some perturbative evaluations of the damping
rate and the Debye screening length.
Choosing a vanishing gauge-fixing parameter (Landau gauge) or introducing
an unphysical infrared cutoff enables us to avoid this gauge-fixing
parameter dependent modification at the long distance introduced
by averaging over a class of gauge conditions.
If the theory is gauge invariant in the way that it can be quantized
by using different gauge constrains (without involving any average),
we can then extract physics from gauge variant propagators evaluated in
a general covariant gauge by choosing a vanishing gauge-fixing parameter.


\acknowledgements

I am grateful to P.~Arnold, S.~Khlebnikov, and especially,
T.~Clark and S.~Love for many helpful discussions.
I also would like to thank L.~S.~Brown and T.~K.~Kuo
for encouragement and advise.
This work was supported by the U.S. Department of Energy,
grant DE-FG02-91ER40681 (Task B).


\newpage

\appendix

\section{Evaluation of $I(x)$ for thermal gauge theory}
For a compactified spacetime, the spacetime point $x$ is understood
as $x = (\tau, {\bf x})$.
Using the dimensional regularization to regulate the ultraviolet
divergence, we evaluate the integral $I(x)$ defined by Eq.~(\ref{I_mom_exp})
as
\begin{eqnarray}
    I (\tau, {\bf x}) &=& T \sum_{k_0}
	\int {d^{d-1} k \over (2 \pi)^{d-1}}
	{1 \over ({\bf k}^2 + k_0^2)^2}
	(1 - e^{ik_0 \tau + i {\bf k} \cdot {\bf x}})
\nonumber\\
	&=& T \sum_{k_0{\not =}0} \int {d^{d-1} k \over (2 \pi)^{d-1}}
        {1 \over ({\bf k}^2 + k_0^2)^2} - T \sum_{k_0{\not =}0}
	\int {d^3 k \over (2 \pi)^3}
	{e^{ik_0 \tau + i {\bf k} \cdot {\bf x}}
	\over ({\bf k}^2 + k_0^2)^2}
\nonumber\\
	&& \qquad + T \int {d^{d-1} k \over (2 \pi)^{d-1}}
	{1 \over k^4} (1 - e^{i {\bf k} \cdot {\bf x}})
	+ O (d-4)
\nonumber\\
	&=& 2 \, T {\Gamma({5-d \over 2})
	\over (4 \pi)^{(d-1)/2}} (2 \pi T)^{d-5} \zeta (5-d)
	- T \sum_{k_0{\not =}0} {1 \over 8 \pi |k_0|}
	e^{i k_0 \tau - |k_0| |{\bf x}|}
	+ { T |{\bf x}| \over 8 \pi} + O (d-4)
\nonumber\\
	&=& {1 \over (4 \pi)^2} (4 \pi T^2)^{d/2-2}
	\left [{2 \over 4-d} + \gamma \right ]
\nonumber\\
	&& \qquad + {1 \over (4 \pi)^2} \left [
	\ln (1 - e^{-2\pi T |{\bf x}| + i 2 \pi T \tau} )
	+ \ln (1 -  e^{-2\pi T |{\bf x}| - i 2 \pi T \tau }) \right ]
	+ { T |{\bf x}| \over 8 \pi} + O (d-4)
\nonumber\\
	&=& {1 \over (4 \pi)^2} (4 \pi T^2)^{d/2-2}
	\left [{2 \over 4-d} + \gamma \right ]
\nonumber\\
	&& \qquad +
	{1 \over (4 \pi)^2} \ln \left (1 + e^{-4 \pi T |{\bf x}|}
	-2 \cos (2 \pi T \tau) e^{-2 \pi T |{\bf x}|} \right )
	+ {T |{\bf x}| \over 8 \pi} + O (d-4) \,.
\label{sumresult}
\end{eqnarray}

\section{Alternate derivation based on the proper vertices Ward identities}
Consider the fermion self energy diagrams in QED. The only place where
the $\xi$ parameter can enter is in the longitudinal part of the photon
propagator $\xi k_\mu k_\nu / k^4$. Therefore, the derivative of
the self energy $\Sigma (p)$ with respect to $\xi$ is
\begin{eqnarray}
    {d \over d \xi} \Sigma (p) &=& q^2 \int (dk) {k_\mu k_\nu \over k^4}
	\Gamma_\mu (k, p, p+k) G(p+k) \Gamma_\nu (-k, p+k, p)
\nonumber\\
	&& + {q^2 \over 2} \int (dk) {k_\mu k_\nu \over k^4}
	\Gamma_{\mu\nu} (k, -k, p, p) \,,
\label{diagrammatic}
\end{eqnarray}
where $\Gamma_\mu (k, p, p+k)$ and $\Gamma_{\mu\nu} (k, k', p, p+k+k')$
are the proper photon-fermion three-point and four-point vertices
respectively. For the proper vertices, our convention is that
the last two momentum arguments are the momentum of
fermion legs while the $k$ and $k'$ denote the photon momenta.
Ward identities\cite{Braaten&Pisarski,Taylor} give
\begin{eqnarray}
    k_\mu \Gamma_\mu (k, p, p+k) &=& G^{-1} (p+k) - G^{-1} (p)
\nonumber\\
    k_\mu k_\nu \Gamma_{\mu\nu} (k, -k, p, p)
	&=& G^{-1} (p+k) + G^{-1} (p-k) - 2 G^{-1} (p) \,.
\end{eqnarray}
Inserting these identities into Eq.~(\ref{diagrammatic}) produces%
\footnote{This equation agrees with the equations appearing in
references\cite{Baier&Kunstatter&Schiff,Rebhan2} where the gauge-fixing
parameter dependence was examined.}
\begin{equation}
    {d \over d \xi} \Sigma (p) = - q^2 G^{-1} (p)
	\int (dk) {1 \over k^4} \left [G (p+k) G^{-1} (p)
	- 1 \right ]
\end{equation}
Since $G(p) G^{-1} (p) = 1$, taking the derivative with respective to $\xi$
yields
\begin{equation}
    {d \over d \xi} \Sigma (p) = {d \over d \xi} G^{-1} (p)
	= - G^{-1} (p) {d \over d \xi} G (p) G^{-1} (p) \,.
\end{equation}
Combining the two equations above, we obtain
\begin{equation}
    {d \over d \xi} G(p) = q^2 \int (dk) {1 \over k^4}
	\left [ G(p+k) - G (p) \right ] \,,
\label{compare}
\end{equation}
which can be written in coordinate space as
\begin{equation}
    {d \over d \xi} G (x)
	= q^2 \int (dk) {1 \over k^4} (e^{ikx} -1) G(x) \,.
\end{equation}
It is straight forward to solve the differential equation above to get
\begin{equation}
    G_\xi (x) = \exp \left \{-\xi q^2
	\int (dk) {1 \over k^4} (1 - e^{ikx}) \right \}
	G_{\xi=0} (x) \,.
\end{equation}
This is the result~(\ref{QEDresult}) derived by functional methods
in the main text.

\end{document}